\documentclass[aps,pra]{revtex4}
\usepackage{graphicx}

\begin{document}

\title{Multiparty Quantum Secret Sharing Based on Entanglement Swapping
\thanks{Email: zhangzj@wipm.ac.cn}}

\author{Zhan-jun Zhang$^{1,2}$ and Zhong-xiao Man$^1$\\
{\normalsize $^1$ Wuhan Institute of Physics and Mathematics,
Chinese Academy of Sciences, Wuhan 430071, China} \\
{\normalsize $^2$ School of Physics \& Material Science, Anhui University, Hefei 230039, China} \\
{\normalsize Email: zhangzj@wipm.ac.cn}}

\date{\today}
\maketitle

\begin{minipage}{410pt}
A multiparty quantum secret sharing (QSS) protocol is proposed by
using swapping quantum entanglement of Bell states. The secret
messages are imposed on Bell states by local unitary operations.
The secret messages are split into several parts and each part is
distributed to a party so that no action of a subset of all the
parties but their entire cooperation is able to read out the
secret messages. In addition, the dense coding is used in this
protocol to achieve a high efficiency. The security of the present
multiparty QSS against eavesdropping has been analyzed and
confirmed
even in a noisy quantum channel.\\

\end{minipage}\\

Suppose Alice wants to send a secret message to two distant
parties, Bob and Charlie. One of them, Bob or Charlie, is not
entirely trusted by Alice, and she knows that if the two guys
coexist, the honest one will keep the dishonest one from doing any
damages. Instead of giving the total secret messages to any one of
them, it may be desirable for Alice to split the secret messages
into two encrypted parts and send each one a part so that no one
alone is sufficient to obtain the whole original information but
they collaborate. To gain this end classical cryptography can use
a technique called as secret sharing [1,2], where secret messages
are distributed among $N$ users in such a way that only by
combining their pieces of information can the $N$ users recover
the secret messages. Recently this concept has been generalized to
quantum scenario [3]. The quantum secret sharing (QSS) is likely
to play a key role in protecting secret quantum information, e.g.,
in secure operations of distributed quantum computation, sharing
difficult-to-construct ancilla states and joint sharing of quantum
money [6], and so on. Hence, after the pioneering QSS work
proposed by using three-particle and four-particle GHZ states [3],
this kind of works on QSS attracted a great deal of attentions in
both theoretical and experimental aspects [4-13,24-26], and
various methods were proposed to realize QSS. Entanglement
swapping [14,15,27] is a method that enables one to entangle two
quantum systems that do not have direct interaction with one
another. Based on entanglement swapping, a number of applications
in quantum information [16] have been found such as constructing a
quantum telephone exchange, speeding up the distribution of
entanglement, correcting errors in Bell states, preparing
entangled states of a higher number of particles, and secret
sharing of classical information. Entanglement swapping is also
used in QSS protocols [7,12], however, in those multi-party QSS
protocols [3,4,11,12] the identification of multi-qubit GHZ states
are required and should be achieved. In fact, according to the
present-day technologies an identification of a Bell state is much
easier than an identification of a GHZ state. In this paper, we
propose a multi-party quantum secret sharing (QSS) protocol based
completely on the entanglement swapping and identification of Bell
states.

Before giving our protocol, let us briefly introduce the local unitary
operations which can impose secret messages on Bell states and the entanglement swapping
of Bell states. Define the four Bell states as
\begin{eqnarray}
|\Psi ^{+}\rangle =\frac{1}{\sqrt{2}}(|0\rangle |1\rangle
+|1\rangle |0\rangle )=\frac{1}{\sqrt{2}}(|+\rangle |+\rangle
-|-\rangle |-\rangle ),\\
|\Psi ^{-}\rangle=\frac{1}{\sqrt{2}}(|0\rangle |1\rangle
-|1\rangle |0\rangle )=\frac{1}{\sqrt{2}}(|+\rangle |-\rangle
-|-\rangle |+\rangle ),\\
|\Phi ^{+}\rangle =\frac{1}{\sqrt{2}}(|0\rangle |0\rangle
+|1\rangle |1\rangle =\frac{1}{\sqrt{2}}(|+\rangle |+\rangle
+|-\rangle |-\rangle ),\\
|\Phi ^{-}\rangle =\frac{1}{\sqrt{2}}(|0\rangle |0\rangle
-|1\rangle |1\rangle )=\frac{1}{\sqrt{2}}(|+\rangle |-\rangle
+|-\rangle |+\rangle ),
\end{eqnarray}
where $|+\rangle=\frac{1}{\sqrt{2}}(|0\rangle+|1\rangle)$ and
$|-\rangle=\frac{1}{\sqrt{2}}(|0\rangle-|1\rangle).$ Let
$u_{1}=|0\rangle\langle0|+|1\rangle\langle1|$,
$u_{2}=|0\rangle\langle0|-|1\rangle\langle1|$,
$u_{3}=|1\rangle\langle0|+|0\rangle\langle1|$,
$u_{4}=|0\rangle\langle1|-|1\rangle\langle0|$ be four local
unitary operators acting on one qubit of the qubit pair in a Bell
state, then one can see that
$u_{1}|\Psi^{-}\rangle=|\Psi^{-}\rangle,
u_{2}|\Psi^{-}\rangle=|\Psi^{+}\rangle,
u_{3}|\Psi^{-}\rangle=|\Phi^{+}\rangle,
u_{4}|\Psi^{-}\rangle=|\Phi^{-}\rangle$. Assume that each of the
above four unitary operations corresponds two classical bits
respectively, i.e., $u_{0}$ to '00', $u_{1}$ to '01', $u_{2}$ to
'10' and $u_{3}$ to '11', then the encodings of the secret
messages can be imposed on the Bell states by using the local
unitary operations. Since the following equations hold,
\begin{eqnarray}
(u_{1}|\Psi^{-}_{ab}\rangle)\otimes|\Psi^{-}_{cd}\rangle=
|\Psi^{-}_{ab}\rangle\otimes|\Psi^{-}_{cd}\rangle=\frac{1}{2}(|\Psi^{-}_{ac}\rangle
|\Psi^{-}_{bd}\rangle+|\Phi^{+}_{ac}\rangle
|\Phi^{+}_{bd}\rangle-|\Psi^{+}_{ac}\rangle
|\Psi^{+}_{bd}\rangle-|\Phi^{-}_{ac}\rangle
|\Phi^{-}_{bd}\rangle),\\
(u_{2}|\Psi^{-}_{ab}\rangle)\otimes|\Psi^{-}_{cd}\rangle=
|\Psi^{+}_{ab}\rangle\otimes|\Psi^{-}_{cd}\rangle=\frac{1}{2}(|\Psi^{+}_{ac}\rangle
|\Psi^{-}_{bd}\rangle-|\Psi^{-}_{ac}\rangle
|\Psi^{+}_{bd}\rangle-|\Phi^{+}_{ac}\rangle
|\Phi^{-}_{bd}\rangle+|\Phi^{-}_{ac}\rangle
|\Phi^{+}_{bd}\rangle),\\
(u_{3}|\Psi^{-}_{ab}\rangle)\otimes|\Psi^{-}_{cd}\rangle=
|\Phi^{+}_{ab}\rangle\otimes|\Psi^{-}_{cd}\rangle=\frac{1}{2}(|\Phi^{-}_{ac}\rangle
|\Psi^{+}_{bd}\rangle-|\Psi^{+}_{ac}\rangle
|\Phi^{-}_{bd}\rangle-|\Psi^{-}_{ac}\rangle
|\Phi^{+}_{bd}\rangle+|\Phi^{+}_{ac}\rangle
|\Psi^{-}_{bd}\rangle),\\
(u_{4}|\Psi^{-}_{ab}\rangle)\otimes|\Psi^{-}_{cd}\rangle=
|\Phi^{-}_{ab}\rangle\otimes|\Psi^{-}_{cd}\rangle=\frac{1}{2}(|\Phi^{+}_{ac}\rangle
|\Psi^{+}_{bd}\rangle+|\Phi^{-}_{ac}\rangle
|\Psi^{-}_{bd}\rangle-|\Psi^{+}_{ac}\rangle
|\Phi^{+}_{bd}\rangle-|\Psi^{-}_{ac}\rangle
|\Phi^{-}_{bd}\rangle),
\end{eqnarray}
obviously, one can see that there is an explicit correspondence between a
known initial state of two qubit pairs (secret encoding has
been imposed on one pair via a local unitary operation) and
its Bell-state measurement outcomes after the
quantum entanglement swapping.

\begin{figure}
\begin{center}
\includegraphics[width=1.0\textwidth]{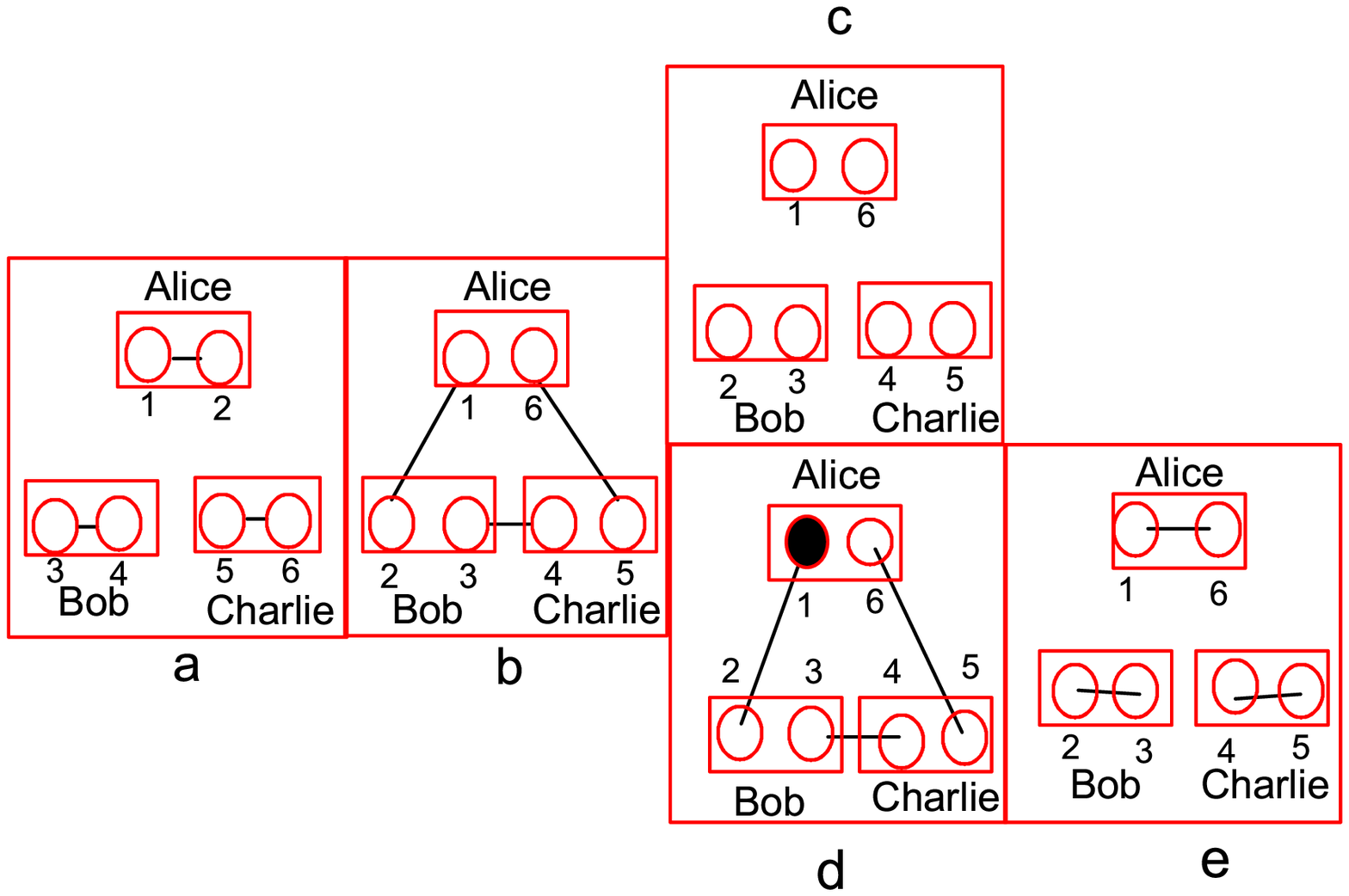}
\vskip -11cm \caption{The detecting mode (a-b-c) and the message
mode (a-b-d-e) of the present quantum secret sharing protocol. The
hollow circle stands for a qubit. The line between two qubits
represents their entanglement. The solid circle in (d) means that
a unitary operation has been performed on the qubit. See text for
detail.} \label{fig1}
 \end{center}
\end{figure}

For convenience, let us first describe a three-party QSS protocol.
Suppose there are three parties, say, Alice, Bob and Charlie. The
sender Alice wants to distribute secret messages between two
parties, Bob and Charlie. To reach this goal, they do as follows.

(S1) Each party prepares two qubits in the same Bell state, say,
$|\Psi ^{-}\rangle$, that is,  Alice (Bob, Charlie) prepares
$|\Psi ^{-}_{12}\rangle$ ($|\Psi ^{-}_{34}\rangle$, $|\Psi
^{-}_{56}\rangle$) (cf., fig1a). Then each party stores one qubit
in its own site and sends another to the specific partner, e.g.,
Alice (Bob, Charlie) sends the qubit 2 (4, 6) to Bob (Charlie,
Alice) (cf., fig1b). They should publicly confirm whether the
success of the qubit distributions has been achieved. If succeed,
Alice can decide to select which one out of the following two
possible choices. With probability $c$ Alice selects the first
choice which we call as {\it detecting mode} hereafter. The aim of
this choice is to check the security of qubit transmission quantum
channels. If this mode is selected, the procedure continues to
(S2). In contrast, Alice can decide to select the second choice
with probability $r=1-c$. The aim of the second choice is to
impose the secret message and implement the QSS. We call this
choice as {\it message mode}. If this mode is selected, the
procedure goes to (S3).

(S2) Alice chooses randomly one of the two sets of measurement
basis (MB), say, $\chi_{z}$= $\{|1\rangle,|0\rangle\}$ and
$\chi_{x}$= $\{|+\rangle,|-\rangle\}$ to measure qubit 1. Then
Alice tells Bob which MB she has chosen and her measurement
outcome. Bob uses the same MB as Alice to measure qubit 2 and
compares his outcome with Alice's (cf., fig1c). If no
eavesdropping exits, their outcomes should be completely opposite,
i.e., if Alice gets $|0\rangle$ $ (|1\rangle)$, then Bob gets
$|1\rangle$ $(|0\rangle)$ and if Alice gets $|+\rangle$
$(|-\rangle)$, then Bob gets $|-\rangle$ $(|+\rangle)$. This
method is sufficient to check whether the Alice-Bob channel is
secure. In fact, in the present protocol there are three qubit
transmission quantum channels, namely, the Alice-Bob, the
Bob-Charlie and the Charlie-Alice quantum channels. Above we only
consider the security of the Alice-Bob channel. Due to symmetry,
security considerations of other channels are same. For
simplicity, here we do not depict others anymore. Only when they
ascertain that there is no Eve in each channel, they turn to (S1).
Otherwise, the QSS is aborted.

(S3) Firstly, Alice performs a local unitary operation randomly on
one of her two qubits 1 and 6 (cf., fig1d). Then she performs a
Bell-state measurement on the qubits 1 and 6 and announces
publicly her measurement outcome. After this, Bob and Charlie
perform Bell-state measurements on their own qubits respectively
and record the measurement outcomes. As a matter of fact, after
Alice's Bell-state measurement, the qubits 2 and 5 should project
to one of the four Bell states (cf., fig1e). If Bob and Charlie
collaborate, according to their Bell-state measurement outcomes
and Alice's public announcement of the Bell-state measurement on
the qubits 1 and 6, they can deduce the exact local unitary
operation which Alice performed on one of her qubits in terms of
eqs.(5-8) in a recursion way. For an example, if Bob's and
Charlie's outcomes are respectively $|\Psi^{-}_{23}\rangle$ and
$|\Phi^{+}_{45}\rangle$, since the state Bob prepared in his
initial qubits 3 and 4 is $|\Psi ^{-}_{34}\rangle$, then from Eq.
(7) they can know that the qubits 2 and 5 has projected to
$|\Phi^{+}_{25}\rangle$ after Alice's Bell-state measurement on
qubits 1 and 6. Since both the initial states of the qubit pair
(1, 2) and the initial states of the qubit pair (5,6) are  $|\Psi
^{-}\rangle$, respectively, and Bob and Charlie have known Alice's
Bell-state measurement outcome on the qubits 1 and 6 (say,
$|\Psi^{+}_{16}\rangle$) and they have already deduced out the
state $|\Phi^{+}_{25}\rangle$ of qubits 2 and 5, then from Eq. (8)
they can know that the local unitary operation performed by Alice
is $u_{4}$, that is, the secret messages Alice distributed is the
two classical bits $'11'$.

So far we have presented a three-party QSS protocol completely
based on the quantum entanglement swapping and identification of
Bell states. Now let us analyze the security of the protocol.
Since for each qubit pair only one qubit is transmitted via a
quantum channel, Eve can not distinguish this Bell state of the
pair with any local operations on this qubit. In order to acquire
Alice's transmitted information, the efficient eavesdropping is to
capture the travel qubits and replace them with their own qubits
prepared previously. But this eavesdropping can be detected in the
detecting mode by using randomly chosen MB and comparing the
measurement outcomes.  Even if in a serious case that an {\it
insider}, say Charlie (Charlie*), cooperate with an outside
eavesdropper Eve, the eavesdropping can also be detected in the
detecting mode. Our protocol is based on EPR pairs, so the proof
of the security is same in essence as those in Ref.[17-21]. Hence
the present protocol is secure against eavesdropping.

Above we have presented a three-party QSS protocol based on
entanglement swapping. In fact, it is easily generalized to a
multiparty case. Suppose there are $N$ parties. At first, each
party prepares two qubits in the Bell state $|\Psi^{-}\rangle$.
Then each of them sends one qubit to the specific partner and
retains another in its own site, that is, the $n$th party prepares
a qubit pair in $|\Psi^{-}\rangle$, then he (or she) sends one
qubit to the ($n+1$)th party and stores one in own site (the $N$th
party sends one qubit to the first party). After this procedure is
successfully finished, they also have two choices. One choice is
to detect eavesdropping. Its deatiled procedure is very similar to
and the same in essence as that in the three-party QSS protocol.
Hence the security of the generalized version can be confirmed.
The other is to distribute the secret messages among the other
parties. The sender (say, Alice, whose $n$ order is assumed to be
the smallest or the largest one) performs a local unitary
operation on one of her two qubits. Then Alice measures this two
qubits in the Bell basis and announces the measurement outcome.
After this, according to the order of $n$ is always increased (or
decreased), each of the other parties performs in turn the
Bell-state measurement on the two qubits in its own site. If they
collaborate, they can successfully extract Alice's secret messages
in a recursive way. Incidentally, in the generalized protocol, the
order of measurement is very important. Once such an order is
destroyed, then the secret message can not be correctly extracted
by the other parties though they collaborate.

It should be pointed out that the above protocol seems to be only
designed for ideal quantum channels. In the above protocol the
reliable sharing of an entangled qubit pair between two parties is
very important and necessary. It is known that when a qubit of an
entangled pair travels in a noisy quantum channel, the initial
entanglement might be lost. Hence the security problem of the
above protocol in a noisy channel seems to arise. Fortunately, it
has been proven that over any long distance two party can reliably
share an entangled pair in terms of the quantum repeater technique
containing the entanglement purification and teleportation[28-32].
Once two parties have shared an entangled qubit pair, then in the
detecting mode any eavesdropping can be detected by using the
method of two MBs. Hence, even in a noisy channel the present
protocol works securely also.

Our protocol owns two distinct advantages over those protocols
using directly multi-particle GHZ states. First, as mentioned in
[12], in the present protocol the parties can apply the
entanglement purification protocol to reliably share a qubit pair
in a Bell state[22,23]. However, for those protocol using GHZ
states, when the number of all the parties is large, how to
prepare a multi-qubit GHZ state and how to reliably share the GHZ
states among multiparties are worthy to be studied further so far
[33]. Secondly, in the present protocol only Bell states are used.
The advantage of such limitation is dominant. For instance, as for
as a ten-party protocol is concerned, if multi-particle GHZ states
are used, one should prepare 511 different multi-particle GHZ
states in advance [See Ref.12] and perform a more difficult
multi-particle GHZ state measurement. However, in the present
protocol, we only need ten Bell states as well as the Bell state
identification. Incidentally, we realize that the experimental
realization of full Bell measurement still represents an unsolved
problem, which affects the advantage over some GHZ-based
protocols.

To summarize, we have presented a multi-party QSS protocol based
on entanglement swapping of Bell states. The security of the
protocol has been confirmed, even in a noisy quantum channel. The
advantages of the present protocol are revealed. \\

\noindent {\bf Acknowledgements}

This work is supported by the National Natural
Science Foundation of China under Grant No. 10304022. \\

\noindent {\bf References}

\noindent[1]  B. Schneier, Applied Cryptography (Wiley, New York,
1996) p. 70.

\noindent[2] J. Gruska,  Foundations of Computing (Thomson
Computer Press, London, 1997) p. 504.

\noindent[3] M. Hillery, V. Buzk  and A. Berthiaume, Phys. Rev. A
{\bf 59}, 1829 (1999).

\noindent[4] R. Cleve, D. Gottesman  and H. K. Lo H K, Phys. Rev.
Lett. {\bf 82}, 648 (1999).

\noindent[5] S. Bandyopadhyay, Phys. Rev. A {\bf 62}, 012308
(2000).

\noindent[6] D. Gottesman,  Phys. Rev. A {\bf 61}, 042311 (1999).

\noindent[7] A. Karlsson, M. Koashi and N. Imoto, Phys. Rev. A
{\bf 59} (1999) 162.

\noindent[8] H. F. Chau, Phys. Rev. A {\bf 66}, 060302 (2003).

\noindent[9] S. Bagherinezhad and V. Karimipour, Phys. Rev. A {\bf
67}, 044302 (2003).

\noindent[10] G. P. Guo and G. C. Guo, Phys. Lett. A {\bf 310},
247 (2003).

\noindent[11] L. Xiao, G. L. Long, F. G. Deng and  J. W. Pan,
Phys. Rev. A {\bf 69}, 052307 (2004).

\noindent[12] Y. M. Li, K. S. Zhang and K. C. Peng, Phys. Lett. A
{\bf 324}, 420 (2004).

\noindent[13] W. Tittel, H. Zbinden and N. Gisin, Phys. Rev. A
{\bf 63} 042301 (2001).

\noindent[14] M. Zukowski, A. Zeilinger, M. A. Horne and A. K.
Ekert, Phys. Rev. Lett. {\bf 71}, 4287 (1993).

\noindent[15] J. W. Pan, D. Bouwmeester, H. Weinfurter and A.
Zeilinger, Phys. Rev. Lett. {\bf 80}, 3891 (1998).

\noindent[16] S. Bose, V. Vedral and P. L. Knight,  Phys. Rev. A
{\bf 57}, 822 (1998).

\noindent[17] F. G. Deng, G. L. Long and  X. S. Liu, Phys. Rev. A
{\bf68},  042317 (2003).

\noindent[18] C. H. Bennett, G. Brassard and N. D. Mermin, Phys.
Rev. Lett. {\bf68},  557 (1992).

\noindent[19] H. Inamori, L. Rallan  and V. Verdral, J. Phy. A
{\bf 34},  6913 (2001).

\noindent[20] G. L. Long and X. S. Liu, Phys. Rev. A {\bf 65},
032302 (2002).

\noindent[21] E. Waks, A. Zeevi and Y. Yamamoto, Phys. Rev. A {\bf
65}, 052310 (2002).

\noindent[22] D. Deutsch  et al, Phys. Rev. Lett. {\bf 77},  2818
(1996).

\noindent[23] B. S. Shi, Y. K. Jiang and G. C. Guo, Phys. Rev. A
{\bf 62}, 054301 (2000).

\noindent[24] V. Scarani and N. Gisin, Phys. Rev. Lett. {\bf 87},
117901 (2001).

\noindent[25] V. Scarani and N. Gisin, Phys. Rev. A {\bf 63},
 042301 (2001).

\noindent[26] A. M. Lance, T. Symul, W. P. Bowen, B. C. Sanders
and P. K. Lam,  Phys. Rev. Lett. {\bf 92}, 177903  (2004).

\noindent[27] Juhui Lee,  Soojoon Lee, Jaewan Kim and Sung Dahm
Oh, Phys. Rev. A {\bf70}, 032305 (2004) .

\noindent[28] C. H. Bennett, D. P. DiVincenzo, J. A. Smolin, W. K.
Wotters, Phys. Rev. A {\bf 54}, 3824 (1996).

\noindent[29] C. H. Bennett, G. Brassard C. Crepeau,  R. Jozsa, A.
Peres and W. K. Wotters, Phys. Rev. Lett. {\bf70}, 1895 (1993).

\noindent[30] H. J. Briegel, W. Dur, J. I. Cirac and P. Zoller,
Phys. Rev. Lett., {\bf 81}, 5932 (1998).

\noindent[31] W. Dur, H. J. Briegel, J. I. Cirac and P. Zoller,
Phys. Rev. A, {\bf 59}, 169 (1998).

\noindent[32] H. K. Lo and H. F. Chau, Science,  {\bf 283}, 2050
(1999).

\noindent[33] Z. Zhao, Y. A. Chen, A. N. Zhang, T. Yang, H. J.
Briegel and J. W. Pan, Nature (London) {\bf 430}, 54 (2004).

\enddocument